\documentclass[aps,prb,superscriptaddress,showpacs,notitlepage,twocolumn,longbibliography]{revtex4-2}
\usepackage{color}
\usepackage{graphicx}
\usepackage[american]{babel}
\usepackage[T1]{fontenc}
\usepackage{amsmath}
\usepackage{amssymb}
\usepackage{amstext}
\usepackage{amsthm}
\usepackage{latexsym}
\usepackage{verbatim}
\usepackage{natbib}
\usepackage{array}
\usepackage{color} 
\usepackage{bm} % bold math, specifically for Greek letters here
\def\Ef{$E_{F}$ }
\def\kz{$k_{z}$ }
\def\dxz{$d_{xz}$}
\def\dyz{$d_{yz}$}
\def\dxy{$d_{xy}$}
\def\dx2my2{$d_{x^2-y^2}$}
\def\dz2{$d_{z^2}$}

\usepackage[draft]{todonotes}   % notes showed

\usepackage{xspace}

\begin{document}

\title {Unfolding the kagome lattice to improve understanding of ARPES in CoSn}

\author{V\'{e}ronique Brouet, Aaditya Vedant}
\affiliation {Universit\'{e} Paris-Saclay, CNRS, Laboratoire de Physique des Solides, 91405, Orsay, France.}

\author{Fran\c cois Bertran}
\affiliation {Synchrotron SOLEIL, L'Orme des Merisiers, D\'epartementale 128, 91190 Saint-Aubin, France}

\author{Patrick Le F\`evre}
\affiliation {Synchrotron SOLEIL, L'Orme des Merisiers, D\'epartementale 128, 91190 Saint-Aubin, France}
\affiliation {Univ Rennes, CNRS, IPR - UMR 6251, F-35000 Rennes, France}

\author{Oleg Rubel}
\affiliation{Department of Materials Science and Engineering, McMaster University, 1280 Main Street West, Hamilton, Ontario L8S 4L8, Canada}

\begin{abstract}
Metallic kagome lattices are attracting significant attention as they provide a platform to explore the interplay between topology and magnetism. Angle-resolved photoemission spectroscopy (ARPES) plays a key role in unraveling their electronic structure. However, the analysis is often challenging due to the presence of multiple bands near the Fermi level. Indeed, each orbital generates three bands in a kagome lattice due to its three sites motif, which soon becomes complicated if many orbitals are present. To address this complexity, using ARPES matrix elements can be highly beneficial. First, band symmetry can be determined through selection rules based on light polarization. We emphasize that, in kagome lattices, as in all multi-site lattices, symmetry of the Bloch state is not only determined by the orbital character but also by the relative phase between the three sublattices. Additionally, interference between the three sublattices leads to a strong modulation of ARPES intensity across neighboring Brillouin zones. We show how unfolded band calculations capture these modulations, helping with band identification. We apply these ideas to CoSn, whose simple structure retains the key features of a kagome lattice. Using polarization dependent ARPES in several Brillouin zones, we isolate the dispersion of each band and discuss novel correlation effects, selectively renormalizing the bands crossing the Fermi level and shifting the others.
	
\end{abstract}

\date{\today}

\maketitle

\section{Introduction}
Metallic compounds with kagome lattices offer an exceptional means to study the interplay between a topological band structure, inherent to the kagome lattice \cite{GuoPRB09,MazinNatCom14}, and magnetism, frequent in 3d kagome systems \cite{YinHasanReview22,WangReview24}. Interesting properties of magnetic kagome materials include a large anomalous Hall effect in the magnetic Weyl semimetal Co$_3$Sn$_2$S$_2$ \cite{LiuNatPhys18}, a Chern quantum phase in TbMn$_6$Sn$_6$ \cite{YinNature20}, and possibly large pseudomagnetic fields in strained FeSn \cite{ZhangNanoLetters23}. Other electronic instabilities, such as exotic charge density waves in the non-magnetic AV$_3$Sb$_5$ \cite{GuoNatPhys24} and magnetic FeGe \cite{TengNature22}, are also under intense investigation. ARPES plays a crucial role in characterizing the electronic structure of these materials \cite{YeNature18,KangNatCom20,KangNatureMat20,LinPRB20,LiuNatCom20,HuReview23,PengCondMat24}, allowing location of Dirac cones, van Hove singularities, and flat bands, characteristic of the kagome band structure. Moreover, it is well suited to characterize the strength of electronic correlations, appearing, for example, as renormalization of the calculated band dispersion. Most metallic kagomes are found to be weakly correlated, with renormalization factors less than 2, but some exhibit more correlations, like CsCr$_3$Sb$_5$ \cite{PengCondMat24}. In these intermetallic kagome systems, a problem is often that the weak crystal field does not significantly lift the degeneracy of the five $3d$ orbitals. This leads to a rather complex situation with 15 bands often strongly overlapping near the Fermi level (5 per $3d$ orbital times the 3 kagome sites).

Fortunately, there are some ways to separate bands in an ARPES experiment. The most common method is to use selection rules associated with linear polarization to select orbitals that are either odd or even with respect to mirror planes containing high symmetry directions \cite{DamascelliRMP03,MoserJElecSpec2023}. In single-site systems, this depends solely on the orbital parity with respect to that mirror plane. In multi-site systems, such as the kagome lattice, this also depends on the relative phases of the orbitals at different sites \cite{BrouetPRB12}, which is not always properly taken into account in the ARPES literature. We will detail these selection rules for a kagome lattice.

\begin{figure}[tbp]
	\includegraphics[width=0.85\linewidth]{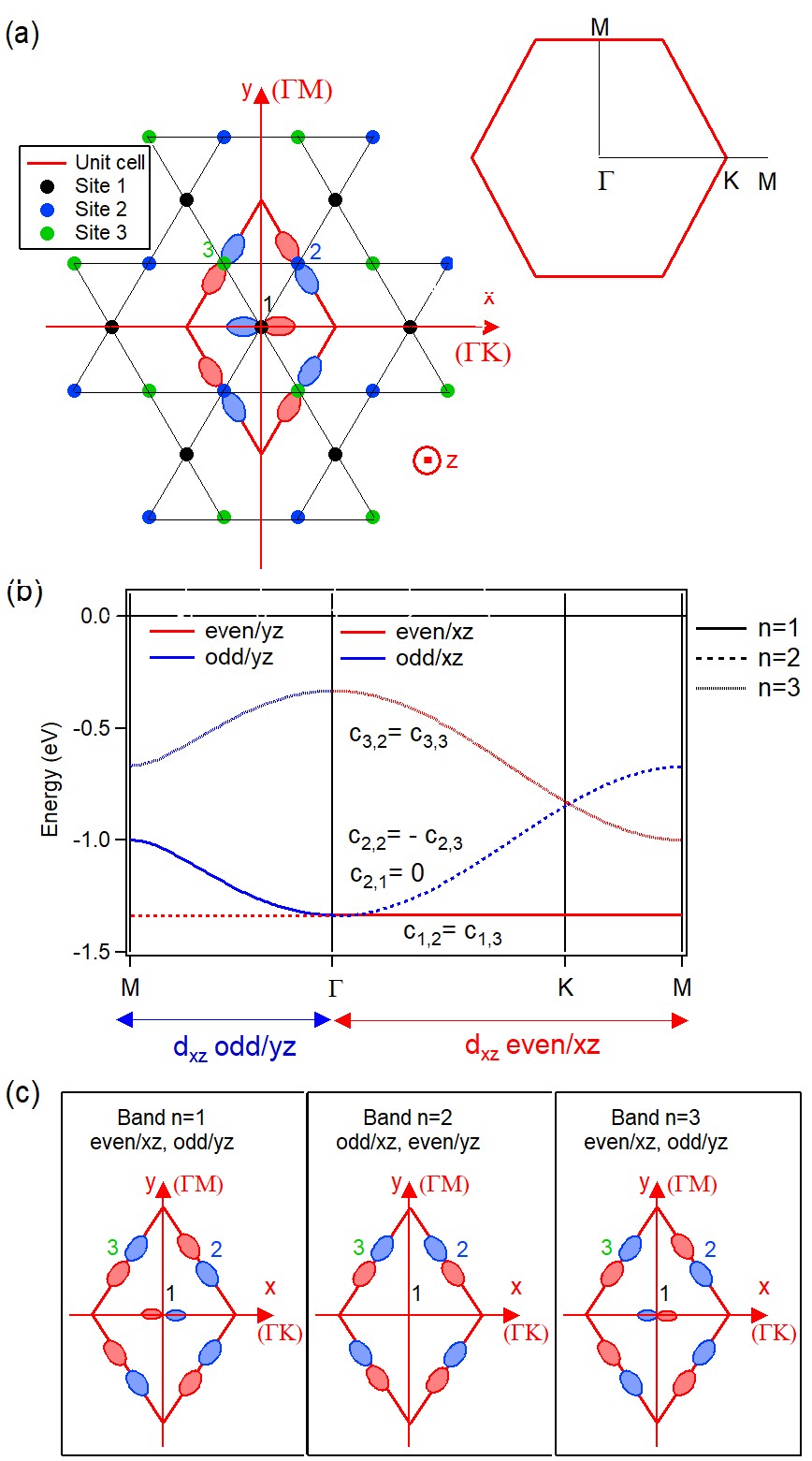}
	\caption{ (a) Sketch of a kagome plane, where the unit cell of parameter a (red line) contains 3 atoms at positions $\bm{\delta}_1=(0,0)$, $\bm{\delta}_{2,3}=(\pm a/4, a\sqrt{3}/4)$. A \dxz~orbital is sketched, viewed from above, at site 1 and rotated by $\pm$120$^\circ$ at the other sites. Inset: Sketch of the corresponding Brillouin zone, with its $\Gamma$, K and M high symmetry points and the path shown in panel (b) as a thin black line. (b) Tight-binding model of the three bands expected for a kagome plane with one \dxz~orbital per site, without SOC. We used parameters facilitating comparison with Fig. \ref{CoSn_DFT} (a hopping parameter of $t=-0.167$~eV and an on site potential of $\epsilon=-3$~eV).  The constraints on the $c_{n,j}(k)$ TB coefficients are indicated for each case (the line type is different for each $n$ value). The color indicates even (red) or odd (blue) symmetry with respect to the mirror plane perpendicular to the $xy$ kagome plane and containing the considered high symmetry direction (i.e. $xz$ plane for $\Gamma$KM or $yz$ for $\Gamma$M). (c) Sketch of the relative phases of the orbital at each site for each band. The values are k-dependent.}
	\label{TB_dxz}
\end{figure}

\begin{figure*}[tbp]
	\center
	\includegraphics[width=0.7\linewidth]{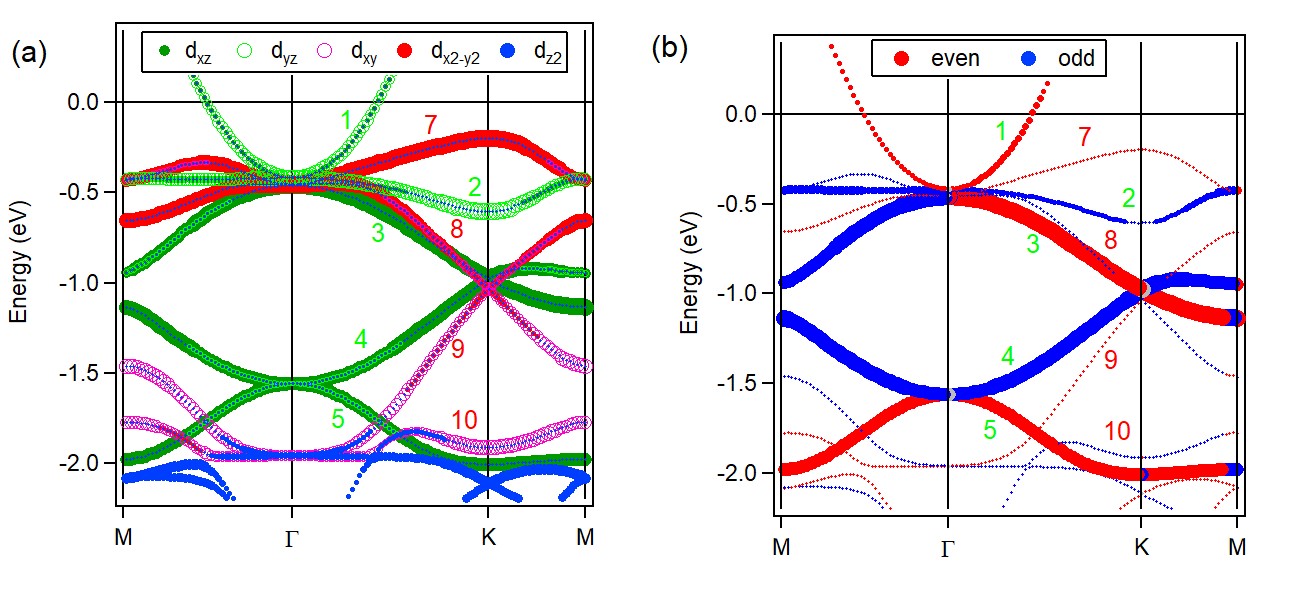}
	\caption{(a) Density functional theory (DFT) calculation for CoSn at \kz=0 without SOC, performed using Wien2k software (similar results were previously obtained \cite{LiuNatCom20}). The orbital characters are coded by different symbols and colors. (b) Same calculation as (a) with \dxz~weight as marker size and parity as color, deduced from the irreducible representation calculated in Wien2k (see text). The parity is given with respect to the mirror plane perpendicular to the surface and containing $\Gamma$M or $\Gamma$KM. }
	\label{CoSn_DFT}
\end{figure*}

Another potentially useful selection rule is the intensity modulation of the bands in different Brillouin Zones (BZ), due to interference between the atoms of the motif. This effect is similar to the form factor in structural experiments, which causes extinctions of some Bragg peaks in multi-site lattices. It naturally appears in ARPES matrix elements \cite{NishimotoJPhysCondMat1996,MoserJElecSpec2023}, but it is much less utilized. For a 2-site motif, there will typically be one bonding and one antibonding band. The interference is constructive only for the bonding band, yielding apparently different dispersion in the first and second BZ. This was first observed in graphite \cite{ShirleyPRB95}, but also in perovskites with rotated octahedra, like iridates \cite{MartinsPRM18} or ruthenates \cite{Sohn2022} and, with some more subtleties, in graphene \cite{MuchaPRB08,LizzitNatPhys10,LiuPRL11} or iron pnictides \cite{BrouetPRB12}. The case of the 3 sites lattice has been much less studied, and it is less predictable. We show here that unfolded calculations \cite{Dirnberger2021,RubelPRB14} can easily predict these effects in complex systems. The relevance of unfolded calculations to analyze ARPES was already indicated for supercell systems \cite{KuPRL10}. More surprisingly, it can also be applied to systems like graphene \cite{LeeJPhysCondMat18}, where the definition of supercell is less intuitive. We show that this can be rationalized within an extended zone scheme. We use here unfolded calculations to obtain the spectral weight of the different bands in different BZ. Although it is not the only contribution to ARPES matrix elements, it often has a significant impact.

We illustrate the benefit of taking these two points (polarization and the use of different BZ) into account with the example of CoSn, a relatively simple  metallic and non-magnetic kagome system. Its electronic structure was already mapped out by ARPES \cite{LiuNatCom20,KangNatCom20,ChengNanoLetters23}, but we pay particular attention to the role of matrix elements. By employing different polarizations along high-symmetry directions, we achieve a complete separation between odd and even bands. We further observe strong modulation of the bands' intensities in different BZs, well predicted by unfolded calculations. We finally discuss the impact of correlations on each band and suggest different behaviors for the bands crossing the Fermi level and the filled ones at higher binding energies. This methodology should be increasingly helpful for more complex kagome systems that deviate more strongly from DFT calculations.

\section{Methods}
Single crystals were grown from a Sn-flux method, as in reference \onlinecite{LiuNatCom20}. Their structure was characterized by X-ray diffraction, their composition by Energy Dispersion X-ray analysis and their electronic properties by transport measurements \cite{sup}. ARPES experiments were carried out on UHV-cleaved single crystals at the CASSIOPEE beamline of SOLEIL synchrotron and the BLOCH beamline of MAX-IV synchrotron, at low temperatures (T$\sim$15~K) and with an overall resolution better than 15meV. 

DFT calculations were performed using the full-potential linearized augmented plane-wave method as implemented in WIEN2k~\cite{blaha2001}, together with the generalized-gradient approximation for the exchange-correlation functional of Perdew, Burke, and Ernzerhof~\cite{Perdew_PRL_77_1996}. The plane-wave expansion in the interstitial region was determined by setting the product $R_{\text{MT}}^{\text{min}}K_{\text{max}}=7$ (the smallest muffin-tin radius and the largest wave vector). The Brillouin zone was sampled using a $\Gamma$-centred $9\times9\times10$ Monkhorst-Pack $k$-mesh, and the tetrahedron method was employed for the integration. Self-consistency convergence criteria were set to $10^{-4}$~Ry for total energy and $10^{-3}$~e for charge. Calculations were carried out using the experimental crystal structure of CoSn (space group $P6/mmm$, No.~191) with lattice parameters $a=b=5.31$~{\AA} and $c=4.24$~{\AA}. The muffin-tin radius was set to 2.49~bohr for Co and Sn.

\section{Band parity in the kagome lattice}

\subsection{Tight-binding model}

%\subsection{B\lowercase{uilding} the 3 bands of the kagome lattice}

The three atoms triangular motif sketched in Fig.~\ref{TB_dxz}(a) is the hallmark of the kagome lattice. For each orbital, three bands $\psi_{n,k}(r)$ will be created, in which the three atoms are combined with coefficients $c_{n,j}(k)$ where j indicates the atomic site at position $\mathbf{\delta_j}$. 
\begin{equation}\label{Eq_TBwave}
	\psi_{n,\mathbf{k}} (\mathbf{r}) = 
	N^{-1/2} \sum_{i,j} c_{n,j}(\mathbf{k})~\mathrm{e}^{\mathrm{i}\mathbf{k}\cdot(\mathbf{R}_i+\bm{\delta}_j)}\, \chi(\mathbf{r}-\mathbf{R}_i-\bm{\delta}_j)
\end{equation}
Let us label the central atom as 1 and the two orthogonal axes as $x$ and $y$. The two planes $xz$ and $yz$, which are perpendicular to the kagome plane, are mirror planes of the structure. They correspond respectively to $\Gamma$K and $\Gamma$M directions of the BZ (see inset in Fig.~\ref{TB_dxz}). One can see that reflection with respect to these mirror planes exchanges atoms 2 and 3 while leaving atom 1 unchanged. Therefore, the three bands created from a single orbital must combine atoms 2 and 3 in a symmetric ($c_{n,2}=c_{n,3}$) or antisymmetric ($c_{n,2}=-c_{n,3}$) way to respect the mirror planes. In the first case, a contribution from atom 1 can be added (in-phase or out-of-phase), while in the second case, it must be zero. A sketch of these combinations for the different values of $n$ is given in Fig.~\ref{TB_dxz}(c).

We take as an example a \dxz~orbital at site 1, which is even with respect to the $xz$ mirror plane and odd with respect to the $yz$ mirror plane. To preserve the symmetry of the kagome lattice, the orbital must rotate by $\pm$120$^\circ$ at sites 2 and 3. This is achieved by taking an appropriate combination of \dxz~and \dyz~orbitals \cite{KimPRB22}. In other words, we consider the even component of the \dxz/\dyz~combination. The typical dispersion of the three bands obtained from a tight-binding (TB) model with only first-neighbor couplings \cite{GuoPRB09} is shown in Fig.~\ref{TB_dxz}(b). We neglect here spin-orbit coupling (SOC) that will open a gap where two bands touch at $\Gamma$ or K \cite{GuoPRB09}. This is crucial to analyse topological properties, but will not play a large role away from the crossing, where we will focus to identify the bands character. We further indicate near each dispersion the relative contributions of the three atoms in the first Brillouin zone and sketch this in Fig.~\ref{TB_dxz}(c). We use color to represent the band parity with respect to mirror planes perpendicular to the kagome plane and containing the high symmetry direction. The parity matches that of the \dxz~orbital when sites 2 and 3 are in phase, and it is opposite when they are out-of-phase. Notably, forming a Dirac cone at K requires bands of opposite parity. However, it is often incorrectly assumed in ARPES studies of kagome compounds that band parity can be inferred directly from the dominant orbital character, leading to a wrong application of the selection rules that we describe below.

\subsection{{CoSn} calculated band structure}

 CoSn has a relatively simple structure with kagome planes stacked vertically on top of each other and separated by two Sn$_2$ buffer layers \cite{SalesPRM21}. This preserves the characteristic features of the simple TB model of the kagome plane, especially Dirac cones and relatively flat bands, as can be seen in the DFT calculation \cite{LiuNatCom20} at $k_z=0$ of Fig. \ref{CoSn_DFT}(a). More precisely, each orbital displays a 3-band structure with very well-defined orbital character (see symbols). We label 1$-$5 the bands built from \dxz/\dyz~orbitals and 7$-$10 bands from \dxy/\dx2my2, observable in the chosen energy window. The \dz2~orbitals appear only below $-2$~eV at \kz=0 and will be ignored for the rest of the discussion (at \kz=$\pi/c$, a strong three-dimensional dispersion brings it much closer to the Fermi level). Two Dirac cones are formed around $-1$~eV for \dxz/\dyz~(bands 3 and 4) and \dxy/\dx2my2 (bands 8 and 9). Two nearly flat bands are present near $-0.4$~eV at $\Gamma$ for \dx2my2 (band 7) and \dyz~(band 2), with a dispersion of about 0.2~eV, much smaller than the 1~eV dispersion of the Dirac bands. 

The three \dxz~bands are clearly visible in this energy window, as highlighted by the marker size in Fig.~\ref{CoSn_DFT}(b). Some residual weight is observed along bands 1 and 2, which are predominantly from \dyz, evidencing hybridization between the two. The symmetry along $\Gamma$M and $\Gamma$K is described by the $C_{2v}$ point group. This symmetry is characterized by two perpendicular mirror planes: the $xy$ kagome plane itself and the perpendicular plane containing the considered symmetry direction. Since the symmetry with respect to this latter mirror plane defines the ARPES selection rules \cite{DamascelliRMP03}, we group in Fig. \ref{CoSn_DFT}(b) the four irreducible representations of $C_{2v}$ into two groups: those that are even (red) and those that are odd (blue) with respect to this mirror plane. One can check that they follow the same rules as \dxz~in the TB model of Fig.~\ref{TB_dxz}(b). 

\subsection{{CoSn} measured band structure with ARPES}

\begin{figure*}[tbp]
	\center
\includegraphics[width=1\linewidth]{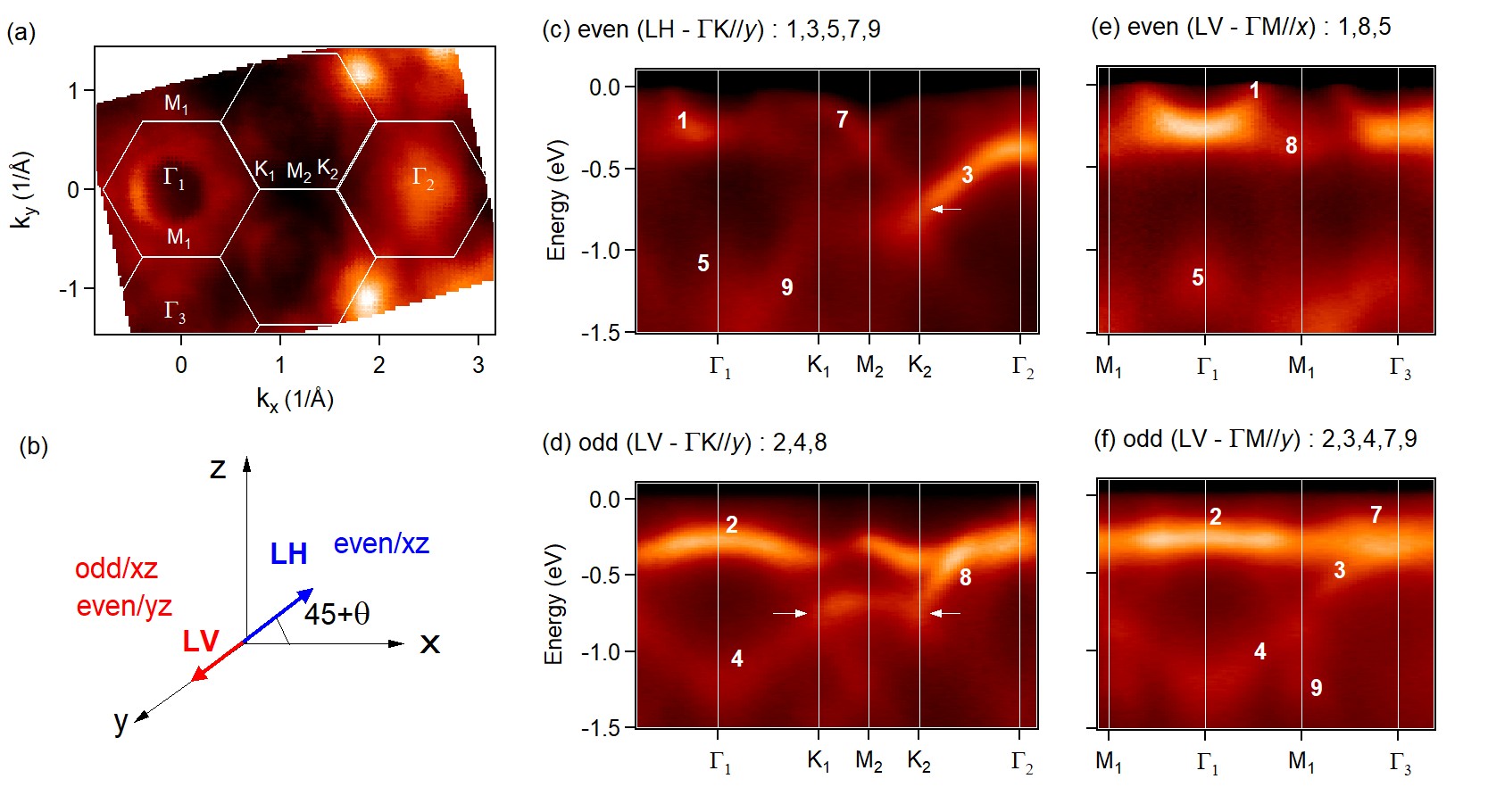}
\caption{(a) Fermi Surface in Co$_{0.95}$Fe$_{0.05}$Sn at 126~eV photon energy and with LH polarization. (b) Sketch of the experimental setup with LH (LV) polarization indicated by blue (red) arrows. The light is incident in the $xz$ plane at an angle depending on $k_x$ fixed by a $\theta$ rotation around $y$. The applicable selection rules are indicated and explained in the text. (c-f) Energy-momentum images along high symmetry directions for even and odd configurations. (c-e) correspond to the orientation of the FS in (a); (f) is rotated by 90$^\circ$. The bands allowed in each configuration are recalled above the corresponding images. The numbers refer to the bands in the calculation of Fig. \ref{CoSn_DFT} to facilitate identification. }
\label{CoSn_ARPES}
\end{figure*}

We now turn to the comparison with actual ARPES experiments. In Fig. \ref{CoSn_ARPES}(a), we show a Fermi Surface for a CoSn sample lightly doped with Fe (5\%), taken with linear horizontal (LH) polarization and at 126~eV photon energy, corresponding to \kz=0 \cite{KangNatCom20}. In ARPES, polarization selection rules apply when the photoelectron momentum $k$ lies in high-symmetry planes, in this case $xz$ and $yz$ containing either $\Gamma$K or $\Gamma$M depending on the sample orientation. If the polarization is even (odd) with respect to this plane, only even (odd) bands will have non-zero matrix elements \cite{DamascelliRMP03,MoserJElecSpec2023}. Fig. \ref{CoSn_ARPES}(b) sketches the experimental geometry with the orientation of LH and LV polarization in our experiment. As LH lies within the $yz$ plane, it is even with respect to it, and even bands will be detected along $y$. As LH has both even and odd components with respect to $xz$, no straightforward selection rules can be applied. On the other hand, LV polarization is along $x$ and perpendicular to $yz$, selecting even bands along $x$ and odd ones along $y$. Note that these selection rules only predict if a band can be observed and not if its intensity will be strong. 

In Fig. \ref{CoSn_ARPES}(c-f), we show measurements along high symmetry directions, where strictly even and odd configurations are realized, either by switching polarization (Fig. \ref{CoSn_ARPES}(d) vs Fig. \ref{CoSn_ARPES}(c), along $\Gamma$K) or the sample orientation (Fig. \ref{CoSn_ARPES}(f) vs Fig. \ref{CoSn_ARPES}(e) for $\Gamma$M). We recall above each image the bands expected to be observed according to parities defined in Fig. \ref{CoSn_DFT}(b). One can check that these selection rules are very well observed with indeed a perfect separation between even and odd bands, as indicated by numbers.

We also observe that the bands are almost systematically observed in only one of the two BZs. At first, this could appear puzzling, as it seemingly breaks the periodicity expected for the kagome lattice. This contrast between different BZs has often been noted in ARPES of kagome systems,but, to our knowledge, it has never been systematically explained.

\section{Unfolding the kagome lattice}

\begin{figure*}[tbp]
	\includegraphics[width=0.9\linewidth]{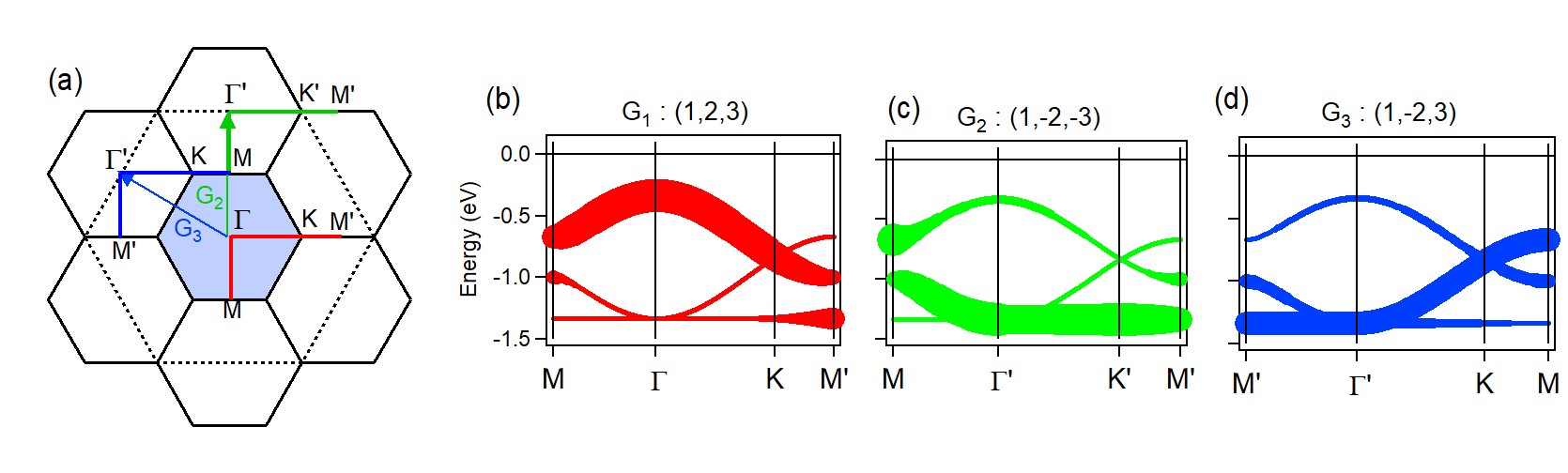}
	\caption{ (a) Sketch of the reciprocal space. The first Brillouin zone (BZ1) is shown in blue, with $\Gamma$, K, and M high symmetry points. The dotted line corresponds to a BZ based on $a/2$, the hidden periodicity of the kagome motif. The points labeled $\Gamma$', K', and M' are inequivalent to those in BZ1 due to a phase factor. (b-d) Weight calculated according to Eq. \ref{TBeq} along the three inequivalent paths shown as red, blue, and green in panel (a). They are translated by the indicated G vectors and the relative phases of the 3 atoms compared to BZ1 are given in parentheses.}    
	\label{TB_dxz_unfolded}
\end{figure*}

\begin{figure*}[tbp]
	\center
	\includegraphics[width=0.8\linewidth]{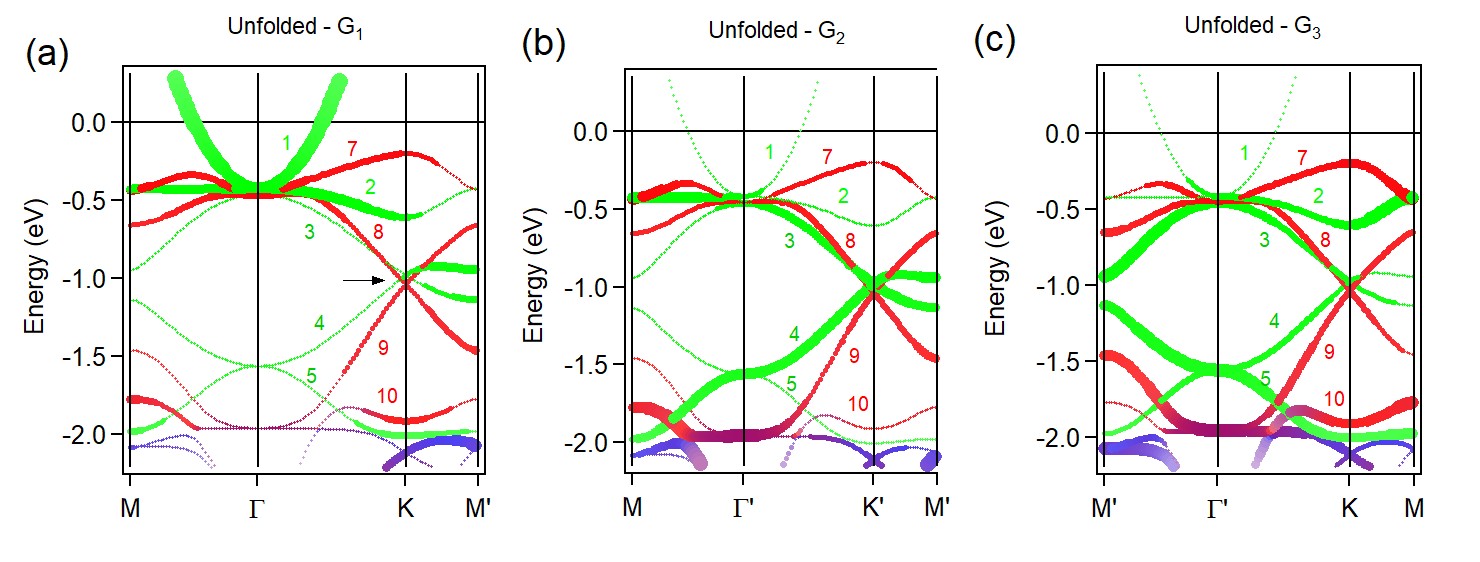}
	\caption{(a-c) Unfolded band structure of CoSn in $k_z=0$ plane indicating the spectral weight as marker size along the three inequivalent M$\Gamma$KM paths (see Fig. \ref{TB_dxz_unfolded}). The color indicates the orbital character: green for \dxz/\dyz, red for \dxy/\dx2my2,~and blue for $d_{z^2}$.}
	\label{CoSn_unfolding}
\end{figure*}

We show here that the expected interferences between the three kagome sites explain this behavior well. We first return to the TB model of the kagome lattice presented in Section~II.A. In an extended zone scheme, we consider the wave vector $\mathbf{q}=\mathbf{k}+\mathbf{G}$, where $\mathbf{k}$ is defined in BZ1 and $\mathbf{G}$ is a reciprocal wave vector of the kagome lattice. From Eq. (1), we see that a phase factor [$\exp(\mathrm{i}~\mathbf{G}\cdot\bm{\delta}_j)$] will appear between values at $\mathbf{k}$ and $\mathbf{q}$, which may be different from one in two neighboring BZ, revealing the hidden periodicity $a/2$ in the motif (see Appendices~\ref{sec:appendix:triangle} and \ref{sec:appendix:extended zone}). A triangular lattice of periodicity $a/2$ would define the larger BZ, represented as a dotted line in Fig.~\ref{TB_dxz_unfolded}(a). In this BZ, there are two inequivalent sets of high-symmetry points, indicated as $\Gamma$, K, M and $\Gamma$', K', M' in the figure. Three inequivalent $M-\Gamma-K-M$ paths are obtained by shifting the path at $\Gamma$ [red, at $\mathbf{G}_1=(0,0)$] by $\mathbf{G}_2=(0,~2\Gamma M)$ (green) or $\mathbf{G}_3=(-3\Gamma K/2,~\Gamma M)$ (blue). $\mathbf{G}_2$ and $\mathbf{G}_3$ are reciprocal wave vectors of the kagome BZ, but not of this larger BZ. In fact, atoms are dephased by $\phi=\mathbf{G}\cdot\bm{\delta}_j$ along these paths, yielding $\pi$ for atoms 2 and 3 along $\mathbf{G}_2$ and $\pi$ for atom 2 only along $\mathbf{G}_3$.

This will directly impact ARPES measurements in different BZs, as it was shown for other multisite systems like graphite or graphene \cite{NishimotoJPhysCondMat1996,Moser16} that the intensity of the ARPES bands is modulated by the sum of the TB coefficients $c_{n,j}(q)$ (see also the Appendix \ref{sec:appendix:matrix elements}). 
\begin{equation}\label{TBeq}
	I_{n}(\mathbf{q})\sim
    \left|
        \sum_{j} c_{n,j}(\mathbf{q})
    \right|^2
\end{equation}
Within the TB framework, this sum can easily be calculated and is represented by marker size in Fig. \ref{TB_dxz_unfolded}(b-d), for each inequivalent path. One can see that only one band is strong for each of the paths, specifically the one with the three atoms in phase (see also Appendix A, which describes the same effect as a function of the hidden periodicities). This emphasizes the need to measure ARPES at least along these three paths to observe all bands. Conversely, this is a means to better identify the nature of the band observed.

\vskip 0.5cm

More generally, when a tight-binding (TB) model is either intractable or fails to capture the relevant physics, the electronic structure can be computed at the DFT level. The resulting wave functions can then be ``unfolded''~\cite{Dirnberger2021,PopescuPRB12,RubelPRB14} by projecting the states obtained in the true unit cell onto the reciprocal space of an alternative unit cell that accounts for additional periodicities $\mathbf{G}_j$
\begin{equation}\label{eq:I(q) unfolding DFT}
	I_{n,\mathbf{k}}(\mathbf{q})
    \propto
    \sum_{\mathbf{G}, \mathbf{G}_j}
    \left|
        C_{n,\mathbf{k}}(\mathbf{G})
    \right|^2
    \delta(\mathbf{k}+\mathbf{G}-\mathbf{q} ~ \mathrm{mod} ~ \mathbf{G}_j)
    .
\end{equation}
Such procedures are now commonly used to compare theoretical calculations with ARPES measurements when a supercell arises from an electronic ordering, such as antiferromagnetism or charge density waves. The resulting folded bands, with periodicity $\mathbf{G}$ imposed by electronic order, typically exhibit weak intensity, in fact proportional to the strength of the perturbation $V(\mathbf{G})$~\cite{VoitScience2000,BrouetPRL04}. This approach can also be beneficial in calculations that involve large unit cells, for example, to simulate doping effects~\cite{BerlijnPRL12}.

We propose here to use it to analyze the kagome lattice, where the origin of the subperiodicities is intrinsic to the multi-site structure (see Appendix \ref{sec:appendix:triangle}). %A similar approach was used for graphene \cite{LeeJPhysCondMat18}. 
To unfold the band structure in the large BZ (dotted line in Fig.~\ref{TB_dxz_unfolded}(a)), we use the \texttt{Fold2Bloch} module \cite{RubelPRB14, Rubel23} and a transformation matrix $\{2~0~0:0~2~0:0~0~1\}$. Figure~\ref{CoSn_unfolding} displays this calculation unfolded along the three different paths identified in Fig.~\ref{TB_dxz_unfolded}(a). The size of the marker gives the unfolded weight corresponding to Eq.~\ref{TBeq} and the colors refer to the orbital character, as in Fig.~\ref{CoSn_DFT}. There are well defined differences expected for spectral weights in different BZ, which will be the largest when all atoms are in-phase. For example band 1 is very strong in (a), while bands 3 and 4 are nearly zero. The opposite is true in panel (b) and (c), but differently for $\Gamma$M and $\Gamma$K. Generally, the unfolding effect is less clear on \dxy/\dx2my2 orbitals, because they are more strongly hybridized with each other. %, blurring the differences in atomic composition (\dxy~and \dx2my2~having opposite parities, they can only be hybridized for different atomic combinations). 

\begin{figure*}[tbp]
	\center
	\includegraphics[width=0.95\linewidth]{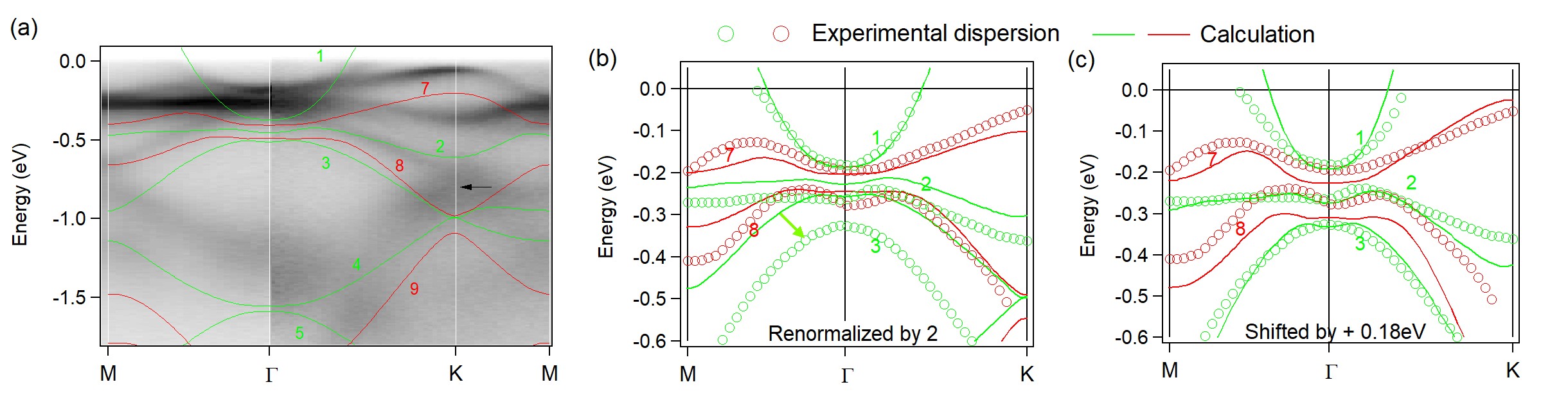}
	\caption{(a) Energy-momentum image of the band structure obtained in CoSn at 126~eV by adding images from ARPES with different polarizations and BZ. The calculated band structure with SOC is overlayed. The black arrow indicates the Dirac cone. (b-c) Experimental models of the bands near \Ef~extracted by tracking the peaks maxima (symbols) and calculation with SOC (lines). Green corresponds to \dxz/\dyz~bands and red to \dxy/\dx2my2. In (b), the DFT calculated eigenvalues are renormalized $[E(k)-E_F]/2$ by a factor of 2, in (c), the DFT calculated eigenvalues are shifted up by 0.18~eV.}
	\label{CoSn_fit}
\end{figure*}

The data in Fig. \ref{CoSn_ARPES}(c-f) should be compared to Fig. \ref{CoSn_unfolding}(a) for BZ1 and Fig. \ref{CoSn_unfolding}(b) for BZ2. Already in the FS, it is clear that the intensity is much larger in BZ1 than in the neighboring BZ2, evidencing these modulations. The circular FS is formed by band 1, the only band crossing $E_F$ at \kz=0. Its larger intensity in BZ1 corresponds very well to prediction of the unfolded calculation. In all dispersion images, there is almost no trace of bands 3 and 8 at $\Gamma$$_1$, as also predicted by unfolding. On the other hand, bands 3 and 8 appear strongly at $\Gamma$$_2$ along $\Gamma$K [Fig. \ref{CoSn_ARPES}(c-d)] and more weakly along $\Gamma$M [Fig. \ref{CoSn_ARPES}(e-f)] in good agreement with Fig. \ref{CoSn_unfolding}(b). Both bands 3 (\dxz/\dyz) and 8 (\dxy/\dx2my2) form Dirac cones at K for similar energies (black arrow), so that separating them as in Fig. \ref{CoSn_ARPES}(c-d) is useful to define their respective dispersions and the energies of the Dirac crossings. On the contrary, band 7, having a flat part near E$_F$ at K, is present in Fig. \ref{CoSn_ARPES}(c) in both BZ, although with stronger weight on slightly different $k$ windows. This also corresponds well to the smaller unfolding effect we noted for \dxy/\dx2my2. The fact that its intensity is distributed over all zones may explain its relatively weaker intensity.

This shows that many strong variations in band intensity across neighboring Brillouin zones (BZs) can be explained by unfolding effects. However, a few deviations are noted compared to the calculated spectral weights. Band 4 appears significantly more intense in BZ1 than in BZ2 [this is particularly evident along the $\Gamma$M path in Fig.~\ref{CoSn_ARPES}(f)], whereas the opposite trend is predicted in Fig.~\ref{CoSn_unfolding}. Band 2 fades away along KM' in Fig.~\ref{CoSn_ARPES}(b), as expected, but becomes very strong along M'K', whereas a strong weight is instead expected along MK in Fig.~\ref{CoSn_unfolding}. These differences may reflect the role of other matrix element effects, which can mask the variation arising from sublattice interference. On one hand, the polarization cross section can vary over $k$-space and, on the other hand, the photon energy cross section can be quite different for different bands.  For example, in CoSn, the intensity of band 7 is weak at 126~eV but much stronger at 70~eV (not shown), while the opposite is true for band 2. More deeply, the ARPES intensity is proportional to the unfolding weight of Eq.~\ref{eq:I(q) unfolding DFT} assuming common approximations (see Appendix C) that can also have their limit.

\section{Electronic correlations in C\lowercase{o}S\lowercase{n}}

Figure~\ref{CoSn_fit}(a) gives an overview of most of the bands, by combining images measured under different experimental conditions in pure CoSn. Five bands (labeled 1$-$3, 7, and 8) are nearly degenerate at $\Gamma$ around $-$0.4~eV, and it is useful to use the combination of unfolding and parity outlined in Fig.~\ref{CoSn_ARPES} to separate them. Although SOC mixes the bands in the crossing region, obscuring the parity and orbital character definitions in this region, they remain  well resolved away from $\Gamma$ for the purpose of this analysis. On the other hand, this mixing is at the origin of the nontrivial topology of CoSn, which was precisely studied in ref.~\onlinecite{KangNatPhys25}. Our findings are in very good agreement with previous ARPES studies of CoSn~\cite{LiuNatCom20,KangNatCom20,ChengNanoLetters23}, which all find a bottom position of band 1 near $-$0.2~eV, the flat band 7 near $-$0.05~eV at $K$ and the Dirac cones around $-$0.8~eV (black arrow). The zoom in the energy range 0.6~eV below \Ef is shown in Fig.~\ref{CoSn_fit}(b,c) with symbols representing the experimental dispersion. Qualitatively, this agrees well with the calculated band structure, including SOC, which is shown as thin lines in Fig.~\ref{CoSn_fit}(a). Quantitatively, however, the calculated dispersions deviate from the experimental data. The band 1 reaches minimum at $-$0.37~eV in calculation, which is significantly different from the experimental value of ca. $-$0.18~eV.

Electronic correlations are expected to renormalize the band dispersions, shrinking the effective band width \cite{DamascelliRMP03}. We apply such a renormalization in Fig.~\ref{CoSn_fit}(b) by dividing all dispersions by a factor of 2 with respect to $E_F$. This correction improves the energy position of band 1 and its curvature, but not the other bands. Band 3 has clearly a higher binding energy than the renormalized calculation (see green arrow). Also, the Dirac cone located experimentally around $-$0.8~eV is now placed at $-$0.5~eV after the factor of 2 renormalization of the calculated band structure. In fact, we observe that the splitting between bands 1 and 3 is 0.1~eV in experiment, exactly as in the unrenormalized band dispersion with SOC. This suggests that this part of the band structure could be well described by simply shifted up the bands. We do this in Fig.~\ref{CoSn_fit}(c), applying an upward shift of 0.18~eV. This describes the dispersions reasonably well within a larger energy window (see band 4 along $\Gamma$M in Fig.~\ref{CoSn_fit}(a), which is also well described by shifting up the calculation), except for band 1. Indeed, such a shift would not conserve the number of charge carriers, contrary to band renormalization that respects the Luttinger theorem~\cite{LuttingerPR60}. As a result, the Fermi crossings $k_F$ of band 1, that was well described by the renormalized band structure, is not reproduced by the shifted one.  %The fact that these crossings correspond well to the bulk calculation, at least at $k_z$=0, suggests that the bulk stoechiometry is conserved.   

A shift is a priori not expected from electronic correlations, but could arise from surface potential \cite{LiuNatCom20,KangNatureMat20,KangNatureMat20_N&V}. %In the related compound FeSn, surface termination dependent surface states were invoked, but this was never considered for CoSn. 
The fact that the bulk stoichiometry is observed here, at least for $k_F$ at $k_z$=0, first suggests that our ARPES data are bulk sensitive. Moreover, surface calculations suggest Sn surface termination for CoSn with no strong deviation from bulk \cite{sup}. We rather suggest here a novel correlation effect, where \textit{only the band crossing the Fermi level is renormalized}, which makes sense as these metallic electrons should be more sensitive to the many-body effects than those of the filled bands. On the other hand, as band 1 is tied by symmetry to other bands (band 1 and band 2 are only separated by the SOC gap, for example), renormalizing band 1 necessarily raises band 2 and the bands strongly hybridized with it. It turns out that the shift corresponding to renormalizing band 1 by a factor of 2 is close to 0.18~eV, which suggests to link the two effects. Further orbital-dependent adjustments may be necessary to improve the description of bands in Fig.~\ref{CoSn_fit}(c), but this reaches the limits of experimental accuracy. We note that a rather large range of renormalization values was given for CoSn, between 1.2 \cite{KangNatCom20,ChengNanoLetters23} and 2 \cite{LiuNatCom20}, probably depending on which features were predominantly fitted. Our hypothesis clarifies why correlations could appear as energy dependent. 

\section{Conclusion}

We have shown how bands can be separated according to their parity through polarization analysis adapted to the kagome lattice and how systematic variation of the ARPES band intensities in neighboring Brillouin zones can be captured by relatively simple unfolded DFT calculations. This provides additional tools to characterize the nature of a band beyond its dispersion. In the non-magnetic kagome compound CoSn, isolating each dispersion suggests that the bands are not identically renormalized, as usually assumed, but that the band crossing the Fermi level is renormalized by a factor of 2, pulling up all other bands by as much as 0.2~eV. This manifestation of electronic correlations appears quite different from what was observed in other multi-orbital systems, such as iron pnictides \cite{YiNPJQuantum17}, where the renormalization rather seems to depend on orbital character. This could be interesting to investigate further.

Following this idea, if many bands cross the Fermi level with different occupied bandwidths, one could expect a rather complex pattern of shifts compared to DFT calculations. In this case, it becomes crucial to have additional ways to identify the nature of a band to make sense of a comparison between ARPES and DFT. This is probably what happens in the magnetic FeSn, where many bands cross the Fermi level and the comparison with DFT calculations is indeed not straightforward at all anymore \cite{KangNatureMat20_N&V} and probably complicated by surface contributions \cite{KangNatureMat20,LinPRB20}. Symmetry-resolved unfolded calculations should help achieve a rational understanding of the ARPES measurement.

\section*{Acknowledgement}
We acknowledge Victor Bal\'edent and the MORPHEUS, MESURES PHYSIQUES and MICRO-NANO platforms of the Laboratoire de Physique des Solides, Universit\'e Paris-Saclay, for help with sample characterization and Alberto Zobelli for introducing us to ref. \onlinecite{LeeJPhysCondMat18}. We acknowledge SOLEIL for provision of synchrotron radiation facilities at the CASSIOPEE beamline (proposal 20221212 and 20230669), the MAX IV Laboratory for beamtime on the BLOCH beamline under proposal 20231202 and Craig Polley and Mats Leandersson for their help during these measurements. Research conducted at MAX IV, a Swedish national user facility, is supported by Vetenskapsrådet (Swedish Research Council, VR) under contract 2018-07152, Vinnova (Swedish Governmental Agency for Innovation Systems) under contract 2018-04969 and Formas under contract 2019-02496. O.R. acknowledges  computing resources provided by the Digital Research Alliance of Canada and travel support from the Natural Sciences and Engineering Research Council of Canada under Discovery Grant RGPIN-2025-05413.

%\vskip 1cm
%%%%%%%%%%%%%%%%%%%%%%%%%%%%%%%%%%%%%%%%%%%%%%%%%%%%%%%%%%%%%%%%%%%%%%%%%%%%%%%%%
\appendix
\section{From triangular to kagome lattice}\label{sec:appendix:triangle}
\begin{figure}[h]
	\includegraphics[width=0.9\linewidth]{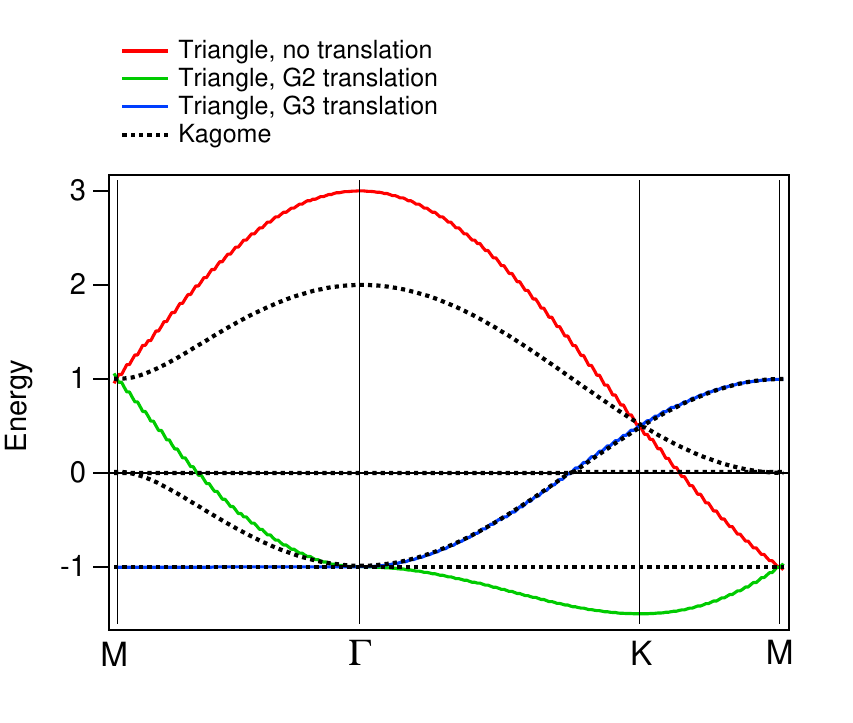}
	\caption{Red line : dispersion of a triangular lattice (unit cell of side a), viewed along the M$\Gamma$KM path of the kagome BZ (unit cell of side 2a, see Fig. \ref{fig_extendedzone}). In green (blue), the dispersion for triangular lattice are folded back with wave vector $\mathbf{G_2}$ ($\mathbf{G_3}$). The black dotted lines are the bands obtained by removing hopping to one of the triangular site. The band for the site disconnected from the others is at E=0. Among the three kagome bands, one is superimposed to the blue line.}   
	\label{Sup_triangle}
\end{figure}

To unfold the kagome lattice, we used the intermediate step of a fictitious triangular lattice. We detail here its relationship with the kagome lattice. 
The dispersion in a tight-binding model for a 2D triangular lattice of side a with nearest-neighbor hoppings has the following form.  

\begin{equation}
E(k_x,k_y)=cos(k_xa)+2cos\left(\frac{k_xa}{2}\right)cos\left(\frac{\sqrt3k_ya}{2}\right)
\end{equation}
In Fig. \ref{Sup_triangle}, we display this dispersion in the kagome unit cell (red line), as well as its translated by $\mathbf{G_2}$=(0,2$\pi$/$\sqrt{3}$a) (green) and $\mathbf{G_3}$=($\pi$/a,$\pi$/$\sqrt{3}$a) (blue), as defined in Fig. \ref{TB_dxz_unfolded}. The resemblance with the kagome dispersion is already apparent. In particular, it is clear that the strongest weight along the three paths indicated in Fig.\ref{TB_dxz_unfolded} follows quite closely the band of the triangular lattice translated by the corresponding wave vector. Let us note that the translation $\mathbf{G_2}\pm\mathbf{G_3}$, also included in an unfolded calculation, is equivalent by symmetry to $\mathbf{G_3}$.  

To obtain the kagome lattice, we "remove" progressively the site at the center of the hexagon by tuning the hopping with this site to zero. We obtain the kagome dispersion shown as dotted black lines, where gap have appeared, where some of the translated bands touched. This is also the regions where the unfolded weight is most distributed between the bands.

\section{Extended zone scheme}\label{sec:appendix:extended zone}

Standard band unfolding techniques rely on a known relationship between the supercell and an assumed primitive cell \cite{Wang_PRL_80_1998, Boykin_71_2005}. Identifying the parent cell is often straightforward in systems with structural distortions (e.g., octahedral tilting~\cite{Rubel23}), magnetic order, or disordered alloys modeled via random atomic substitutions in a supercell~\cite{RubelPRB14}. However, in some cases, superlattice effects are more subtle and difficult to disentangle. A prominent example is quasicrystals, which display sharp Bragg peaks despite lacking conventional long-range periodicity~\cite{Bindi_S_324_2009}. The extended zone scheme provides a way to represent the electronic structure without assuming a specific primitive cell or Brillouin zone. This approach has previously been applied to aperiodic systems to extract a quasi-dispersion relation~\cite{Niizeki_JPCM_2_1990, Mayo_JPCM_32_2020}.

Let us assume that we wish to construct an extended zone representation for an eigenstate $n$ of an aperiodic structure expressed in a tight-binding basis
\begin{equation}\label{eq:appendix:TB wave func (non-periodic)}
	\psi_n (\mathbf{r}) = 
	N^{-1/2} \sum_{j} c_{n,j} \, \chi(\mathbf{r}-\bm{\delta}_j).
\end{equation}
Here, $\bm{\delta}_j$ labels the atomic site positions, $N$ is the total number of sites, $c_{n,j}$ denotes the site amplitude, and $\chi(\mathbf{x})$ is an atom-centered basis function. Due to the lack of translational symmetry, this state does not possess a well-defined crystal momentum. The spectral function $A_n(\mathbf{q})$ in the extended zone scheme is obtained via the generalized structure factor~\cite{Niizeki_JPCM_2_1990}
\begin{equation}\label{eq:appendix:A_n}
	A_n(\mathbf{q}) = N^{-1}
	\left|
	\sum_{j} c_{n,j} \, e^{-\mathrm{i}\mathbf{q} \cdot \bm{\delta}_j}
	\right|^2 .
\end{equation}
This result is analogous to the diffraction pattern of a quasicrystal, where $c_{n,j}$ serves as a site-dependent scattering amplitude~\cite{Niizeki_JPCM_2_1990}.

In periodic solids, an electronic eigenstate with wave vector $\mathbf{k}$ can be expanded in plane waves
\begin{equation}\label{eq:appendix:psi(n,k) plane wave}
	\tilde{\psi}_{n,\mathbf{k}} (\mathbf{r}) 
	= \frac{1}{\sqrt{\Omega}} \sum_{\mathbf{G}} C_{n,{\mathbf{k}}} (\mathbf{G}) \, e^{\mathrm{i} (\mathbf{G}+\mathbf{k}) \cdot \mathbf{r}} ,
\end{equation}
where $\Omega$ is the unit cell volume, and $C_{n,{\mathbf{k}}} (\mathbf{G})$ are the plane wave expansion coefficients. The summation is carried out over reciprocal vectors
\begin{equation}
	\mathbf{G} = m_1 \mathbf{b}_1 + m_2 \mathbf{b}_2 + m_3 \mathbf{b}_3
	\quad \quad
	(m_1, m_2, m_3 \in \mathbb{Z})
\end{equation}
with $\mathbf{b}_i$	denoting the reciprocal lattice basis vectors. By analogy with Eq.~\eqref{eq:appendix:A_n}, the spectral function can be written as
\begin{equation}
	\begin{split}
		A_{n,\mathbf{k}}(\mathbf{q}) &
		= \Omega^{-1}
		\left|
		\int_{\Omega} \sum_{\mathbf{G}} C_{n,{\mathbf{k}}} (\mathbf{G}) \, e^{\mathrm{i} (\mathbf{G}+\mathbf{k}) \cdot \mathbf{r}} \, e^{-\mathrm{i}\mathbf{q} \cdot \mathbf{r}}~\mathrm{d}\mathbf{r}
		\right|^2\\
		& = 
		\left|
		\sum_{\mathbf{G}} C_{n,{\mathbf{k}}} (\mathbf{G}) \, \delta(\mathbf{G}+\mathbf{k} - \mathbf{q})
		\right|^2 .
	\end{split}
\end{equation}
The Dirac delta function effectively selects a single plane wave component, satisfying $\mathbf{q}=\mathbf{G}+\mathbf{k}$, that contributes to the spectral weight. Alternative formulations of the spectral function in the extended zone, aimed at analyzing ARPES measurements, have also been proposed in the literature \cite{Kosugi_JPSJ_86_2017, Mayo_JPCM_32_2020}.

In ARPES, the measured intensity reflects the momentum-resolved spectral function near the surface region. To simulate this, we compute the spectral function projected onto a plane parallel to the surface (001)
\begin{widetext}
    \begin{equation}
    	\bar{A}_{n,\mathbf{k}}(q_x, q_y)
    	= \Omega^{-1}
    	\sum_{\mathbf{G}}
    	\left|
    	C_{n,{\mathbf{k}}} (\mathbf{G})
    	\right|^2 
    	\delta(G_x + k_x - q_x, G_y + k_y - q_y)
    	,
    	\quad
    	\mathbf{k} \in (k_x, k_y, 0)
    	.
    \end{equation}
    The incoherent sum over plane wave coefficients in this expression implies that Fourier components with nonzero out-of-plane wave vectors are folded back into the projection plane. The Fermi surface within the $(q_x, q_y,0)$ plane in the extended zone representation can then be constructed as
    \begin{equation}\label{eq:appendix:A_FS}
    	\bar{A}_{\text{FS}}(q_x, q_y)
    	= 
    	\sum_{n,\mathbf{k}}
    	\bar{A}_{n,\mathbf{k}}(q_x, q_y) \,
    	\delta[E_n(\mathbf{k})-E_{\text{F}}]
    	,
    	\quad
    	\mathbf{k} \in (k_x, k_y, 0)
    	.
    \end{equation}
\end{widetext}

\begin{figure}[h]
	\centerline{\includegraphics[width=0.95\linewidth]{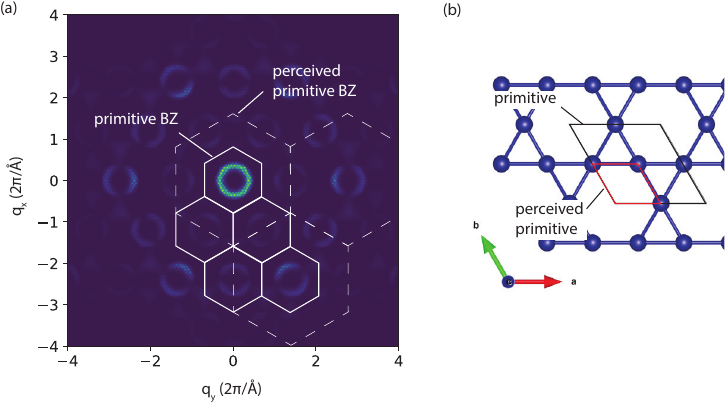}}
	\caption{(a) Extended zone representation of the spectral function corresponding to the Fermi surface of CoSn in the $(q_x, q_y, 0)$ plane. The projected spectral function is computed according to Eq.~\eqref{eq:appendix:A_FS}. The solid line outlines the first Brillouin zone of the primitive cell along with its periodic images. The dashed line indicates a proposed alternative Brillouin zone, which better reflects the observed periodicity in reciprocal space. (b) Real-space lattices corresponding to the primitive cell (black) and the perceived primitive cell (red).}
	\label{fig_extendedzone}
\end{figure}

Figure~\ref{fig_extendedzone}(a) shows a slice through the Fermi surface of CoSn calculated using the Vienna \textit{ab initio} simulation package (VASP)~\cite{Kresse_PRB_47_1993, Kresse_CMS_6_1996, Kresse_PRB_54_1996} and presented in the extended zone scheme. Exchange-correlation effects were described within the generalized-gradient approximation of Perdew, Burke, and Ernzerhof (PBE)~\cite{Perdew_PRL_77_1996}. Projector-augmented-wave potentials (version 5.4)~\cite{Kresse_PRB_59_1999} were employed, with 15 and 4 valence electrons for Co and Sn, respectively. The cutoff energy for the plane-wave expansion was set to 271~eV, corresponding to the maximum \texttt{ENMAX} value among the atomic species in the \texttt{POTCAR} file. A $\Gamma$-centered $7 \times 7 \times 7$ k-mesh was used to sample the BZ. A Gaussian smearing of width 0.01~eV was applied to facilitate BZ integration and ensure smooth convergence of the electronic self-consistent cycle. The electronic energy convergence threshold was set to $10^{-7}$~eV. SOC was included.

A circular pattern appears in the first Brillouin zone but is absent in the adjacent periodic image, reappearing only in the third shell. This doubling of periodicity in reciprocal space suggests a Brillouin zone that is twice as large, which corresponds to a perceived primitive cell that is twice as small in real space, as illustrated in Fig.~\ref{fig_extendedzone}(b). This result supports the validity of the unfolding method and confirms the appropriateness of the chosen primitive cell.

\section{ARPES matrix elements}\label{sec:appendix:matrix elements}

Comprehensive treatments of ARPES principles can be found in textbooks \cite{HufnerBook13,MalterreBook24} and papers \cite{DamascelliRMP03,Moser16,YenCondMat24}. Applying Fermi's golden rule, the photoemission intensity will depend on matrix elements of the form:
\begin{equation}
	I\left(k,\omega\right)
    =
    \frac{2\pi}{\hbar}
    \sum_{i,f}
    \left|
        \left\langle
            f\left|H_{int}\right|i
        \right\rangle
    \right|^2\ \delta(E_f-E_i-\hbar\omega)
\end{equation}

where $H_{int}$ describes the interaction between electron and photon, $|i\rangle$ and $|f\rangle$ are initial and final states. These are usually $N$-particle states, but for our discussion, only the one-electron matrix element will matter. $H_{int}$ is usually developed as a perturbation in the dipolar approximation $H_{int}=\frac{e}{m}\mathbf{A}\cdot\mathbf{p}$, although this is not sufficient for some cases (especially when the Hamiltonian is non-local~\cite{HwangPRB11}). The final state is usually taken as a plane wave of wave-vector $k_F$, assuming the photoelectron is free to travel to the detector. Corrections to this approximation are often important to properly describe the dependence on photon energy or polarization \cite{KernPRR23}. Unfortunately, this often requires sophisticated calculations, which miss a simple connection with the properties of the initial state. 

We consider the matrix element $M_{if}$=$\left\langle f\middle|H_{int}\middle|i\right\rangle$. Using the TB expression given in Eq. \ref{Eq_TBwave}~as initial state, we get.
\begin{equation}
M_{if}=\sum_{i,j}{e^{\mathrm{i}\mathbf{k}\cdot\left(\mathbf{R}_i+\bm{\delta}_j\right)}{\ c_{n,j}\left(\mathbf{k}\right)}\left\langle f\middle|H_{int}\middle|\mathbf{R}_i+\bm{\delta}_j\right\rangle}
\end{equation}
To calculate the ARPES intensity, we follow Nishimoto et al., who derived Eq.~\eqref{TBeq} in the case of graphite \cite{NishimotoJPhysCondMat1996}. A similar derivation can be found elsewhere \cite{Moser16,YenCondMat24}. The term in brackets essentially depends on the shape of the orbital at site $\mathbf{R}_i+\bm{\delta}_j$. They are all the same in our case and we call $M_{orb}$=$\left\langle f\middle|H_{int}|R_0\right\rangle$, where $R_0$ is a lattice site. Recalling the final state $f(\mathbf{r})$ is a Bloch function of wave vector $\mathbf{k}_f$, we have 
$f(\mathbf{r}+\mathbf{R}_i+\bm{\delta}_j)=e^{\mathrm{i}\mathbf{k}_f\cdot(\mathbf{R}_i+\bm{\delta}_j)}f(\mathbf{r})$. 

Therefore,
\begin{equation}
M_{if}=M_{orb}\sum_{i,j}{e^{\mathrm{i}(\mathbf{k}-\mathbf{k}_f)\cdot\mathbf{R}_i}{e^{\mathrm{i}(\mathbf{k}-\mathbf{k}_f)\cdot\bm{\delta}_j}{\ c_{n,j}(\mathbf{k})}}}
\end{equation}
Using $\sum_{i}^{N_{cells}} e^{\mathrm{i}(\mathbf{k}-\mathbf{k}_f)\cdot\mathbf{R}_i}=N_{cells}~\delta(\mathbf{k}-\mathbf{k}_f)$, we finally get:
\begin{equation}
	M_{if}=M_{orb} * \sum_{j}{\ c_{n,j}(\mathbf{k})},
\end{equation}
\noindent which leads to Eq.~\eqref{TBeq}. 

If we rather write the initial state as a function of the sublattice periodicities $|k+G_j>$ (see Appendix A), the delta function selects a value $|k+G_j>$=$k_F$, leading to Eq.~\eqref{eq:I(q) unfolding DFT}.

\bibliography{Biblio_kagome_CoSn,Biblio_Gal}

\end{document}